\documentclass[aps,prd,twocolumn,groupedaddress,nofootinbib,showkeys,showpacs,sort&compress]{revtex4}
\usepackage{graphicx}
\usepackage{amsmath}

\begin{document}

\title{The QCD chiral transition temperature in a Dyson-Schwinger-equation context}

\author{M. Blank}
\email[]{martina.blank@uni-graz.at}
\affiliation{Institut f\"ur Physik, Karl-Franzens-Universit\"at Graz, A-8010 Graz, Austria}

\author{A. Krassnigg}
\email[]{andreas.krassnigg@uni-graz.at}
\affiliation{Institut f\"ur Physik, Karl-Franzens-Universit\"at Graz, A-8010 Graz, Austria}

\date{\today}

\begin{abstract}
We analyze the chiral phase transition with the help of the QCD gap equation. Various
models for the effective interaction in rainbow truncation are contrasted with regard
to the resulting chiral transition temperatures. In particular, we investigate possible
systematic relations of the details of the effective interaction and the value of $T_c$. 
In addition, we quantify changes to the transition temperature beyond the rainbow truncation.
\end{abstract}

%\keywords{}

\maketitle

\section{Introduction\label{intro}}
The structure of the phase diagram of quantum chromodynamics (QCD) has
recently received a lot of attention. Theoretical progress in various
directions is well in line with past, present, and upcoming experimental programs 
e.g.~at CERN SPS, RHIC, LHC, and FAIR. Among the key features of the phase diagram is the
chiral phase transition at nonzero temperature and vanishing chemical potential.
This transition has been and is studied extensively by various different approaches
with an emphasis on the critical behavior of the order parameter as well as
on the actual value of the transition temperature $T_c$.
Lattice QCD calculations for example over the past years yielded $T_c$ at temperatures 
between $145$ and $200$ MeV with a recent tendency towards a commonly found value
\cite{Bernard:2004je,Aoki:2006br,Cheng:2006qk,Aoki:2006we,Cheng:2007jq,Bazavov:2009zn,Aoki:2009sc}.
In various model approaches in mean-field approximation and beyond one finds a range from
$90$ to $160$ MeV, see e.g., \cite{Schaefer:2004en} and references therein.

Although an understanding of the universal properties of QCD is of great importance,
it is also rewarding to investigate the value of $T_c$ which is not 
universal and can therefore be used to draw conclusions about
specific details of the underlying dynamics of quarks and gluons. More precisely,
specific models and their assumptions can be falsified in the approach presented here
by a calculation of $T_c$ in QCD.

\section{Finite-temperature applications of Dyson-Schwinger equations\label{dsesatft}}
A nonperturbative continuum approach to QCD is provided by the Dyson-Schwinger
equations (DSEs) \cite{Fischer:2006ub,Roberts:2007jh}, which can be applied straight-forwardly
also for finite temperature and chemical potential \cite{Roberts:2000aa}. The DSEs are
well-suited to study the properties of QCD's
elementary degrees of freedom, quarks and gluons. In this framework, various aspects
of chiral symmetry and its dynamical breaking have been studied some time ago via the DSE
of the quark propagator (also referred to as the ``QCD gap equation'')
at zero temperature (see, e.\,g.~\cite{Maris:2003vk} and references therein) and in connection with
the phase diagram of QCD (see, e.\,g.~\cite{Roberts:2000aa} and references therein).
The same framework also allows the consistent and covariant study of bound-state, i.\,e.~meson
and baryon, properties. At zero temperature, a level of sophistication beyond simple QCD modeling
has been reached (see e.\,g.~\cite{Krassnigg:2009zh,Eichmann:2007nn,Eichmann:2009qa} and references therein).
At nonzero temperature and density, analogous investigations are complicated by the additional
variables introduced in the Matsubara formalism. Nonetheless, investigations of meson masses at and beyond
the phase transition at zero chemical potential
\cite{Maris:2000ig,Blaschke:2000gd,Blaschke:2000zm,Blaschke:2006ss,Horvatic:2007wu,Horvatic:2007qs,Alkofer:1989vr}
as well as certain aspects of the phase diagram
\cite{Maris:1997eg,Blaschke:1999ab,Nickel:2006vf,Nickel:2006kc,Chen:2008zr,Klahn:2009mb}
have been conducted at various but mostly simpler
degrees of sophistication. In the present context, our focus is on DSE studies of the chiral 
phase transition in QCD at finite temperature and zero chemical potential. 

It was shown some time ago that certain classes of models
in the QCD gap equation in rainbow truncation yield a second-order phase transition with mean-field critical exponents
\cite{Holl:1998qs,Maris:1999bj}. While the main emphasis of that investigation was the critical behavior and
universality class of the phase transition, it also provided the corresponding values for $T_c$,
which ranged between $120$ and $174$ MeV among the model interactions investigated. In somewhat
different setups, using a separable interaction kernel in the gap equation
\cite{Blaschke:2000gd,Blaschke:2006ss,Blaschke:2007ce,Horvatic:2007wu} or neglecting retardation
effects \cite{Alkofer:1989vr}, one obtaines values in the range between $110$ and $146$ MeV.

The pointwise
behavior of all these interactions as a function of the gluon momentum is rather different, although in
all cases it was determined by adjusting the relevant parameters to meson phenomenology at zero
temperature. Therefore, it is interesting to ask whether there is any simple relationship between the
momentum dependence of the interaction and the value of $T_c$. Such a relationship could e.g.~be
analogous to that found recently in studies of meson properties using a particular form
of model interaction with a one-parameter setup. There the value of the free parameter
determines an effective range of the intermediate- and low-momentum parts of the interaction.
While ground-state properties of pseudoscalar and vector mesons were unaffected by variations
in the model parameter, masses of excitations of any kind (orbital or radial) showed a strong and systematic
dependence on the model parameter
\cite{Holl:2004fr,Holl:2004un,Holl:2005vu,Krassnigg:2008gd,Krassnigg:2009zh} thus identifying
excited-state properties as prime targets to study the particular details of the effective interaction
in the DSE formalism, in particular in the nonperturbative regime.
In an analogous fashion, the goal of the present work is to identify $T_c$ as a quantity with
the same capability. In particular, one can expect a calculation of $T_c$ to impose even stronger restrictions
on model parameters, beyond what is at hand at zero temperature.

The paper is organized as follows: Sec.~\ref{formalism} briefly sketches the quark DSE in QCD
at finite temperature, Sec.~\ref{models} compares the model interactions used herein,
Sec.~\ref{transition} outlines how we compute $T_c$ and the critical exponents. The results
are presented in Secs.~\ref{results} and \ref{brt}, the latter focusing on effects from corrections
beyond rainbow truncation. Conclusions and an outlook are presented in Sec.~\ref{conclusions}. All
calculations are performed in Euclidean space.

\section{QCD gap equation at finite temperature\label{formalism}}
We use the quark DSE to study the chiral phase transition of QCD. 
It is an inhomogeneous nonlinear integral equation for the dressed-quark propagator,
whose inverse at $T=0$ is of the form
\begin{equation}\label{eq:propt0}
S^{-1}(p) = i \gamma\cdot p A(p^2) + B(p^2),
\end{equation}
where the four-vector $p$ denotes the quark momentum and $A$ and $B$ are scalar functions of the
momentum-squared. At nonzero temperature the presence of the heat bath implies an additional
four vector, which leads to more structures in (\ref{eq:propt0}) \cite{Kalashnikov:1980bm}.
In the Matsubara formalism one arrives at
\begin{eqnarray}\nonumber
S^{-1}(\vec{p},\omega_k) &=& i \vec{\gamma}\cdot\vec{p}\;A(\vec{p}\,^2,\omega_k) +
B(\vec{p}\,^2,\omega_k) \\\label{eq:prop}
&+&i \gamma_4\omega_k C(\vec{p}\,^2,\omega_k) +
\vec{\gamma}\cdot\vec{p}\;\gamma_4\omega_k  D(\vec{p}\,^2,\omega_k).
\end{eqnarray}
Now the four scalar functions ($A,B,C,D$) depend on the quark 3-momentum squared $\vec{p}\,^2$ and the
quark's fermion Matsubara frequency $\omega_k = 2 \pi T (k + 1/2)$, $k\in\mathbf{Z}$.

The QCD gap equation in pictorial form reads
\begin{equation}\label{eq:gap}
(\parbox[c]{0.12\columnwidth}{\includegraphics[width=0.12\columnwidth]{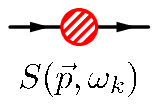}})^{-1} =
(\parbox[c]{0.12\columnwidth}{\includegraphics[width=0.12\columnwidth]{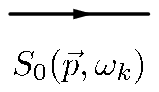}})^{-1} +
\parbox[c]{0.40\columnwidth}{\includegraphics[width=0.40\columnwidth]{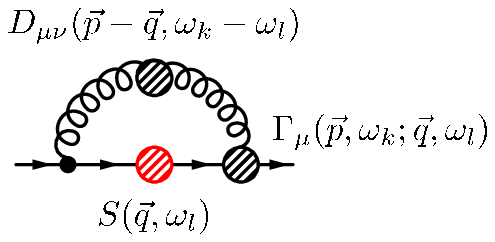}} \;,
\end{equation}
where $S_0(\vec{p},\omega_k)$ is the free quark propagator, $D_{\mu\nu}(\vec{p}-\vec{q},\omega_k-\omega_l)$ the
renormalized dressed gluon propagator and $\Gamma_\mu(\vec{p},\omega_k;\vec{q},\omega_l)$ the
renormalized dressed quark-gluon vertex. Solving for $S(\vec{p},\omega_k)$ one needs expressions for
the quark-gluon vertex as well as the gluon propagator. These can in principle be obtained
\begin{table}
\caption{\label{tab:models}Parameter sets $D$ and $\omega$ used for the different interactions
MN \cite{Munczek:1983dx}, Eq.~(\ref{eq:mn});
AWW \cite{Alkofer:2002bp}, Eq.~(\ref{eq:aww});
MT \cite{Maris:1999nt}, Eq.~(\ref{eq:mt});
MR \cite{Maris:1997tm}, Eq.~(\ref{eq:mr});
together with corresponding observables in $T=0$ meson studies, where available.
Numbers are in GeV except for $D$ (GeV${}^2$) and the chiral condensate,
whose dimension is given explicitly. }
\begin{ruledtabular}
\begin{tabular}{l|c c c c c}
Model&$\omega$&$D$&$-\langle \bar{q}q\rangle_0$&$m_\pi$&$m_\varrho$\\
\hline
MN&&0.5618&    $(.115\, \mbox{GeV})^3$&0.14 &0.77\\
AWW1&0.3&1.47 &$(.245\, \mbox{GeV})^3$&0.135&0.745\\
AWW2&0.4&1.152&$(.246\, \mbox{GeV})^3$&0.135&0.748\\
AWW3&0.5&1.0&  $(.251\, \mbox{GeV})^3$&0.137&0.758\\
MT1&0.3&1.24&  $(.243\, \mbox{GeV})^3$&0.139&0.747\\
MT2&0.4&0.93&  $(.242\, \mbox{GeV})^3$&0.139&0.743\\
MT3&0.5&0.744& $(.239\, \mbox{GeV})^3$&0.141&0.724\\
MR1&0.3&0.78&  $(.241\, \mbox{GeV})^3$&0.139&\\
MR2&0.4&0.78&  $(.250\, \mbox{GeV})^3$&0.139&\\
MR3&0.5&0.78&  $(.255\, \mbox{GeV})^3$&0.139&
\end{tabular}
\end{ruledtabular}
\end{table}
from their own DSEs self-consistently, which is clearly beyond the scope of the present study. Instead we
use various model Ans\"atze in a rainbow truncation of the gap equation, as specified below.
Defining the gluon four momentum as $(\vec{k},\Omega):=(\vec{p}-\vec{q},\omega_k-\omega_l)$, the
second term on the right-hand side of (\ref{eq:gap}), the quark self energy $\Sigma(\vec{p},\omega_k)$, is
given by
\begin{eqnarray}\label{selfenergy}
\Sigma(\vec{p},\omega_k) & =& T\!\!\sum_{l=-\infty}^\infty\!\! \int \frac{d^3q}{(2\pi)^3}
\frac{4}{3} g^2 D_{\mu\nu}(\vec{k},\Omega)\times\\
&&\gamma_\nu S(\vec{q},\omega_l)\Gamma_\mu(\vec{p},\omega_k;\vec{q},\omega_l)  .
\label{regself}
\end{eqnarray}
The gap equation is subject to the renormalization condition that for a renormalization
scale $\zeta$ and $\vec{p}\,^2+\omega_0^2=\zeta^2$
\begin{equation}
S^{-1}(\vec{p},\omega_0) =i\vec{\gamma}\cdot\vec{p} + i\gamma_4\omega_0 + m(\zeta).
\end{equation}
Note that this condition does not contain a term $\vec{\gamma}\cdot\vec{p}\;\gamma_4\omega_k$, i.e.,
the scalar function $D(\vec{p}\,^2,\omega_k)$ will be power-law suppresed in the ultra-violet \cite{Roberts:2000aa}.

In the following, unless otherwise noted, we use the rainbow-truncated gap equation by setting
$\Gamma_\mu(\vec{p},\omega_k;\vec{q},\omega_l) = \gamma_\mu$. The finite-$T$ expression for
the renormalized dressed gluon propagator in Landau gauge reads \cite{Kapusta:1989tk,Roberts:2000aa}
\begin{eqnarray}\nonumber
g^2 D_{\mu\nu}(\vec{k},\Omega) &=&
P_{\mu\nu}^L(\vec{k},\Omega)\;\mathcal{G}(\vec{k},\Omega;m_g) + \\ \label{eq:gluonprop}
&& P_{\mu\nu}^T(\vec{k},\Omega)\; \mathcal{G}(\vec{k},\Omega;0) ,\\
P_{\mu\nu}^T(\vec{k},\Omega)&=&
\left\{\begin{matrix}0,& \mu\;\mbox{and/or}\;\nu=4\\
\delta_{ij}-\frac{k_i k_j}{\vec{k}^2},&\mu,\nu=i,j=1,2,3\end{matrix}\right.,\\
P_{\mu\nu}^L(\vec{k},\Omega)&=&\delta_{\mu\nu}-\frac{k_\mu k_\nu}{\vec{k}^2+\Omega^2}-P_{\mu\nu}^T,
\end{eqnarray}
where $m_g$ is a Debye mass \cite{Bender:1996bm} and $\mathcal{G}$ is an effective interaction detailed below.

\section{Interaction Models}\label{models}
At zero temperature covariant bound-state equations have been used
to study the properties of hadrons in various particular setups in order to gain insight
regarding the properties of the underlying fundamental Green functions of QCD from a
phenomenological point of view. More concretely, for the case of mesons the Bethe-Salpeter-Equation
in ladder truncation together with the gap equation in rainbow truncation has been employed using
many different Ans\"atze for the quark-gluon interaction to investigate various meson properties, see \cite{Krassnigg:2009zh} and
references therein. In that process, different effective interactions have been tested by employing
model-independent constraints for the interaction and fixing the remaining (in our case either one or two)
parameters and a current-quark mass to phenomenological and experimental numbers such as the chiral condensate,
the pion mass, and the pion decay constant.

In the present work we investigate four different functional forms of the interaction, referred to in the following by
MN (Munczek-Nemirovsky \cite{Munczek:1983dx}), AWW (Alkofer-Watson-Weigel \cite{Alkofer:2002bp}),
MT (Maris-Tandy \cite{Maris:1999nt}), and MR (Maris-Roberts \cite{Maris:1997tm}), in order of complexity.
The zero-temperature effective interaction can be straightforwardly used at finite temperature by
insertion in Eq.~(\ref{eq:gluonprop}). Via $s := \vec{k}\,^2 + \Omega^2 + m_g^2$  one gets
\begin{eqnarray}\nonumber
\mathcal{G}(\vec{k},\Omega;m_g) &=&\\ \label{eq:mn}
\textrm{MN}:&=&
D\;\frac{4 \pi^3}{T} \delta^3(\vec{k}) \delta_{k-l,0}\\ \label{eq:aww}
\textrm{AWW}:&=&
D\; \frac{4 \pi^2}{\omega^6} s e^{-s/\omega^2}\\ \label{eq:mt}
\textrm{MT}:&=&
D\; \frac{4 \pi^2}{\omega^6} s e^{-s/\omega^2} + \mathcal F_{UV}(s)\\ \nonumber
\textrm{MR}:&=&
D\; \left( \frac{4 \pi^3}{T} \delta^3(\vec{k}) \delta_{k-l,0} +
\frac{4 \pi^2}{\omega^6} s e^{-s/\omega^2}\right)\\ \label{eq:mr}
&& + \mathcal F_{UV}(s),\\ \nonumber
\mathcal F_{UV}(s):&=&\frac{4\pi\;\gamma_m \pi\;\mathcal{F}(s) }{1/2 \ln [\tau\!+\!(1\!+\!s/\Lambda_\mathrm{QCD}^2)^2]}
\end{eqnarray}
As given in \cite{Maris:1999nt}, ${\cal F}(s)= [1 - \exp(-s/[4 m_t^2])]/s$, $m_t=0.5$~GeV,
$\tau={\rm e}^2-1$, $N_f=4$, $\Lambda_\mathrm{QCD}^{N_f=4}= 0.234\,{\rm GeV}$, and $\gamma_m=12/(33-2N_f)$.
At $T=0$ all of these Ans\"atze provide the correct amount of dynamical chiral symmetry breaking as well as quark confinement
via the absence of a Lehmann representation for the dressed quark propagator. In addition, MT and MR produce the
correct perturbative limit of the QCD running coupling, i.e.~they preserve the one-loop renormalization-group
behavior of QCD for solutions of the quark DSE. Thus, for MT and MR, a renormalization procedure is needed in
the gap equation, which we implement following Refs.~\cite{Maris:1997tm,Maris:2000ig}, where the necessary
details can be found.

\begin{figure}
%\begin{center}
\includegraphics[scale=1,angle=0,clip=true]{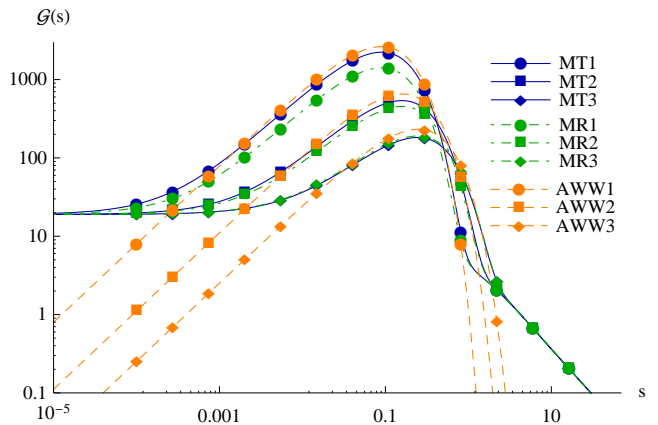}
\caption{(Color online) Plot of the different forms for the effective interaction used in this work
as simple functions of a momentum squared $s$. The corresponding parameters can be found in Tab.~\ref{tab:models}.
Note the different behavior of the interactions on all momentum ranges. In particular, the reader should keep
in mind that the $\delta$-function present in MR, Eq.~(\ref{eq:mr}) is
invisible in the double-logarithmic presentation as well as the MN interaction, Eq.~(\ref{eq:mn}),
which is zero for $s\ne 0$.\label{fig:int}}
%\end{center}
\end{figure}
Here, for the interactions in Eqs.~(\ref{eq:aww}), (\ref{eq:mt}), and (\ref{eq:mr}), we investigate different
parameter sets (i.e., values of $D$ and $\omega$) in analogy to the corresponding bound-state studies at $T=0$.
Table \ref{tab:models} lists these sets of parameters together with the respective results for relevant $T=0$
observables. Once $D$ and $\omega$ have been fitted to the chiral condensate, one remains with a $u/d$
current-quark mass to be chosen in order to calculate properties of light mesons. For each set the
light quark mass has been adjusted to fit $m_\pi$; $m_\varrho$ is listed for a first corresponding
result without fixing any further parameters. Since our study will remain in the chiral limit, the meson
masses are included only to complete the justification of the model parameters.

For easy reference and an instructive comparison, the effective interaction for all parameter sets
is plotted in Fig.~\ref{fig:int}. Differences are clearly visible in the UV regime. In the IR one
has to keep in mind that the $\delta$-function present in the MR interaction, Eq.~(\ref{eq:mr}) is
not shown here due to the double-logarithmic presentation as well as the MN interaction, Eq.~(\ref{eq:mn}),
which is zero for $s\ne 0$. The intermediate
momentum region is dominated by the Gaussian peak whose position and width are characterized by the
parameter $\omega$ and whose overall strength is $D$.

The reader may ask whether there is any relation between $D$ and $\omega$ that could simplify the
parameterization. Indeed, inspection of Tab.~\ref{tab:models} shows that for the MT interaction
the product $\omega D$ is constant over the range of values for $\omega$ used here, when one aims
at a given value of the chiral condensate. In addition, without changing the value of the $u/d$ current-quark
mass, also $m_\pi$ (and approximately also $m_\varrho$) are unchanged on this domain, which can be used
to motivate a redefinition of the parameters \cite{Eichmann:2008ae}. As already mentioned in Sec.~\ref{dsesatft},
one can interpret this pattern with regard to $\omega$ such that this parameter represents an inverse effective
range of the interaction, a dependence on which should be less prominent in ground-state properties than
for excited states. However, for the other interaction types, such a simple relation among the parameters
does not exist and so we keep the interactions in their present form.
\begin{figure}
%\begin{center}
\includegraphics[scale=1,angle=0,clip=true]{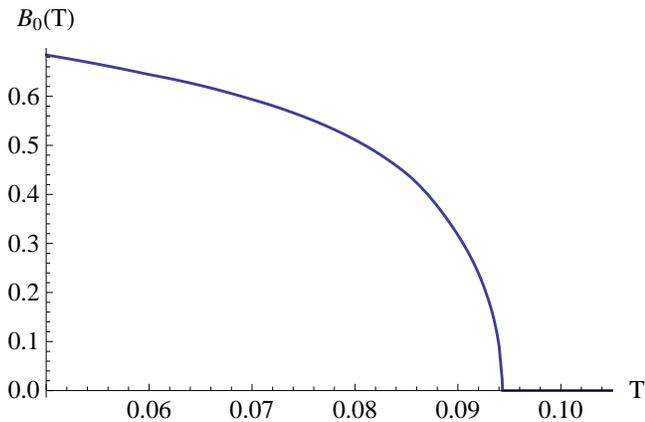}
\caption{(Color online) Plot of the order parameter $B_0$ vs. temperature $T$, shown exemplarily
for model MT2.\label{fig:orderpar}}
%\end{center}
\end{figure}

\section{Chiral Phase Transition}\label{transition}
With all ingredients of the QCD gap equation at hand one can obtain a solution numerically and study the chiral
\begin{figure}
%    \begin{center}
\includegraphics[scale=0.85,angle=0,clip=true]{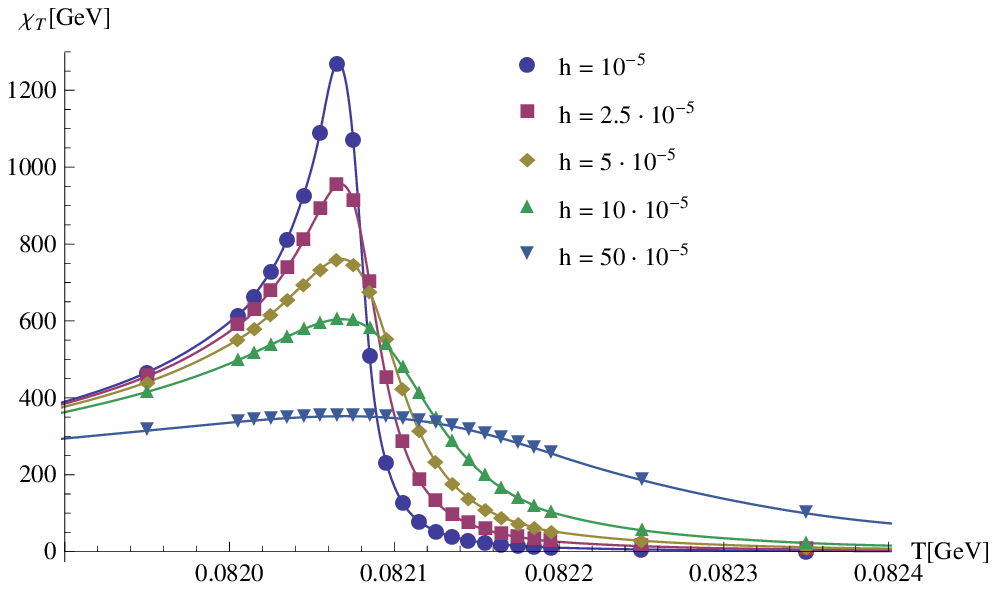}
\includegraphics[scale=0.85,angle=0,clip=true]{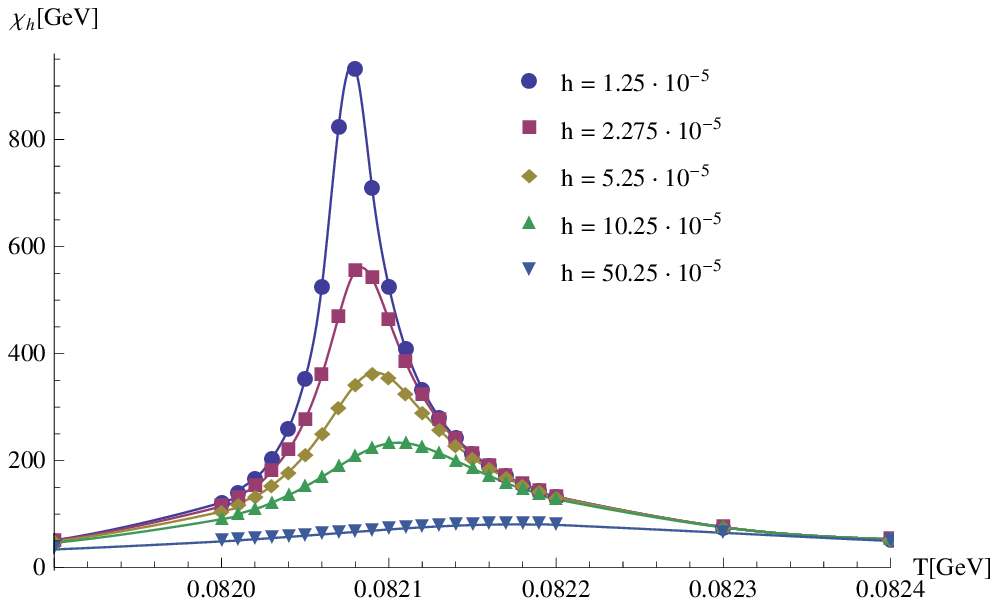}
\caption{(Color online) Chiral susceptibilities $\chi_T(T)$ (upper panel) and $\chi_h(T)$ (lower panel) for the MT1 model
plotted versus temperature $T$ for various values of $h$. The curves between the data points were obtained by
interpolation using cubic splines.\label{fig:pseudocritical}}
%      \end{center}
\end{figure}
phase transition temperature $T_c$ as well as the nature of the transition. The corresponding order
parameter is the chiral condensate. Equivalently, we use $B_0:=B(0,\omega_0)$, since it is easier to access and
can be calculated more accurately. A typical behavior of the order parameter for a chiral-limit solution
is shown in Fig.~\ref{fig:orderpar} and indicates a second order phase transition. Indeed as mentioned earlier, it has been found
previously that the rainbow-truncated quark DSE generally yields a second order chiral phase transition
with mean-field critical exponents \cite{Holl:1998qs,Maris:1999bj}. We have confirmed this behavior for all
interactions considered here and thus enlarged the set of interaction types it had been shown for by MT and AWW.
The following account of our procedure includes the discussion of critical exponents mainly for completeness,
but also, since as described below, $T_c$ can be extracted more accurately this way. In the case of a second
order phase transition, the order parameter $B_0(t,h)$ obeys the scaling laws

\begin{eqnarray}
B_0(t,h) &\propto& \left.(-t)^{\beta}\right|_{h=0}\;\;\;t\rightarrow0^-\quad \text{and}\label{eq:scaling1}\\
B_0(t,h) &\propto& \left.h^{1/\delta}\right|_{t=0}\;\;\;h\rightarrow0^+\;.\label{eq:scaling2}
\end{eqnarray}
Here, $t = \frac{T}{T_c}-1$ is the reduced temperature and $h=\frac{m}{T}$ the reduced mass, a measure
for the explicit breaking of chiral symmetry by a non-vanishing current quark mass $m$. $\beta$ and
$\delta$ are the critical exponents of the phase transition.

Although it is straightforward to use Eqs.~(\ref{eq:scaling1})-(\ref{eq:scaling2}) to obtain $\beta$ and $\delta$,
this procedure requires to solve the gap equation at $h=0$ or $T=T_c$. This is numerically difficult, such that
a direct evaluation does not allow fits to extract $T_c$ with the necessary precision and reliably
observe scaling. Therefore
we exploit further scaling relations and use chiral susceptibilities for our analysis. They are defined by
\begin{eqnarray}
 \chi_T &:=& \left.-T_c \frac{\partial B_0(T,h)}{\partial T}\right|_{h\;\text{fixed}}\;,\quad\text{and}\\
\chi_h &:=& \left.\frac{\partial B_0(T,h)}{\partial h}\right|_{T\;\text{fixed}}\;.
\end{eqnarray}
The maxima of these quantities for nonvanishing $h$ are referred to as pseudocritical points,
$\chi_T^{pc}$ and $\chi_h^{pc}$, respectively. The corresponding pseudocritical temperatures
are denoted by $T_T^{pc}$ and $T_h^{pc}$. They are obtained as the maxima of the chiral susceptibilities with
respect to temperature, as depicted in Fig.~\ref{fig:pseudocritical}.

\begin{figure}
%    \begin{center}
                    \includegraphics[scale=0.95,angle=0,clip=true]{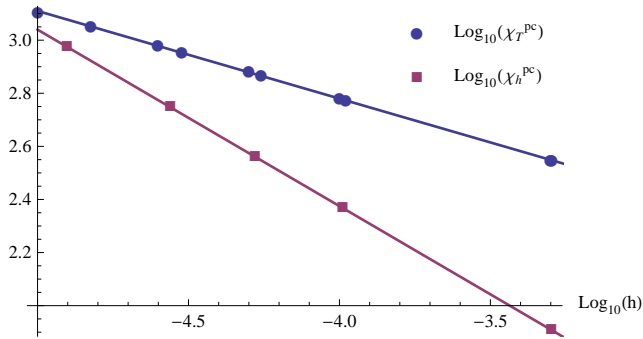}
                    \caption{(Color online) Peak heights of the chiral
susceptibilities $\chi_T^{pc}$ and $\chi_h^{pc}$ plotted versus the reduced mass $h$ on log-log
 scale, for the MT1 model. The lines are linear fits through the calculated points which are used
 to obtain the critical exponents according to Eqs.~(\ref{eq:scaling3}) - (\ref{eq:scaling4}).\label{fig:scaling}}
%       \end{center}
\end{figure}
The pseudocritical points $\chi_T^{pc}$ and $\chi_h^{pc}$ also obey scaling laws. Their behavior for
$T\sim T_c$ and $h\sim 0$ is described by
\begin{eqnarray}
 \chi_T^{pc} &\propto& h^{-1+1/\delta}\;,\label{eq:scaling3}\\
\chi_h^{pc} &\propto& h^{\frac{1}{\beta\delta}\left(1-\beta\right)}\;.\label{eq:scaling4}
\end{eqnarray}
Following \cite{Blaschke:1998mp} we use these scaling relations to obtain $\beta$, $\delta$, and
$T_c$ for the all models and parameter sets given in Tab.~\ref{tab:models}, as illustrated
in Fig.~\ref{fig:scaling}.

\section{Results and discussion\label{results}}
As already mentioned above, all interactions yield a second order phase transition
with mean-field critical exponents, regardless of the strength of the interaction,
its range, or its pointwise behavior in any particular momentum range. In particular,
the presence of the $\delta$-function term does not make a difference in this respect.
Another general observation is the fact that the function $D(\vec{p}\,^2,\omega_k)$
in Eq.~(\ref{eq:prop}) is identically zero for all interactions in rainbow truncation.

The values obtained for the chiral transition temperature $T_c$, however, are rather
different among the various interactions. In the following we will argue how this
feature can be exploited to discern various forms within a given truncation.
The results for $T_c$ are summarized in Tab.~\ref{tab:results} for the four interactions
on a range of model parameters well-used in meson phenomenology.
\begin{table}
\begin{ruledtabular}
\caption{Results for the transition temperature $T_c$ rounded to MeV for the interactions
and parameter sets defined in
Tab.~\ref{tab:models}.\label{tab:results}}
\begin{tabular}{l r r r r }
$\omega$ [GeV] & N/A & 0.3 & 0.4 & 0.5  \\\hline
Model& MN & & &  \\
$T_c$& 169& & &  \\\hline
Model& & AWW1& AWW2& AWW3 \\
$T_c$& &  82 &  94 & 101  \\\hline
Model& &  MT1&  MT2&  MT3\\
$T_c$& &  82 &  94 &  96  \\\hline
Model& &  MR1&  MR2&  MR3\\
$T_c$& & 120 & 133 & 144
\end{tabular}
\end{ruledtabular}
\end{table}

The MN interaction has to be treated somewhat separately, since the only free parameter in
this case is $D$ whose choice completely determines $\mathcal{G}$. In terms of plain numbers
it is interesting to see that MN yields the highest value for $T_c$, followed by MR.
AWW and MT give the smallest numbers.
\begin{figure*}
\includegraphics[scale=0.37,angle=270,clip=true]{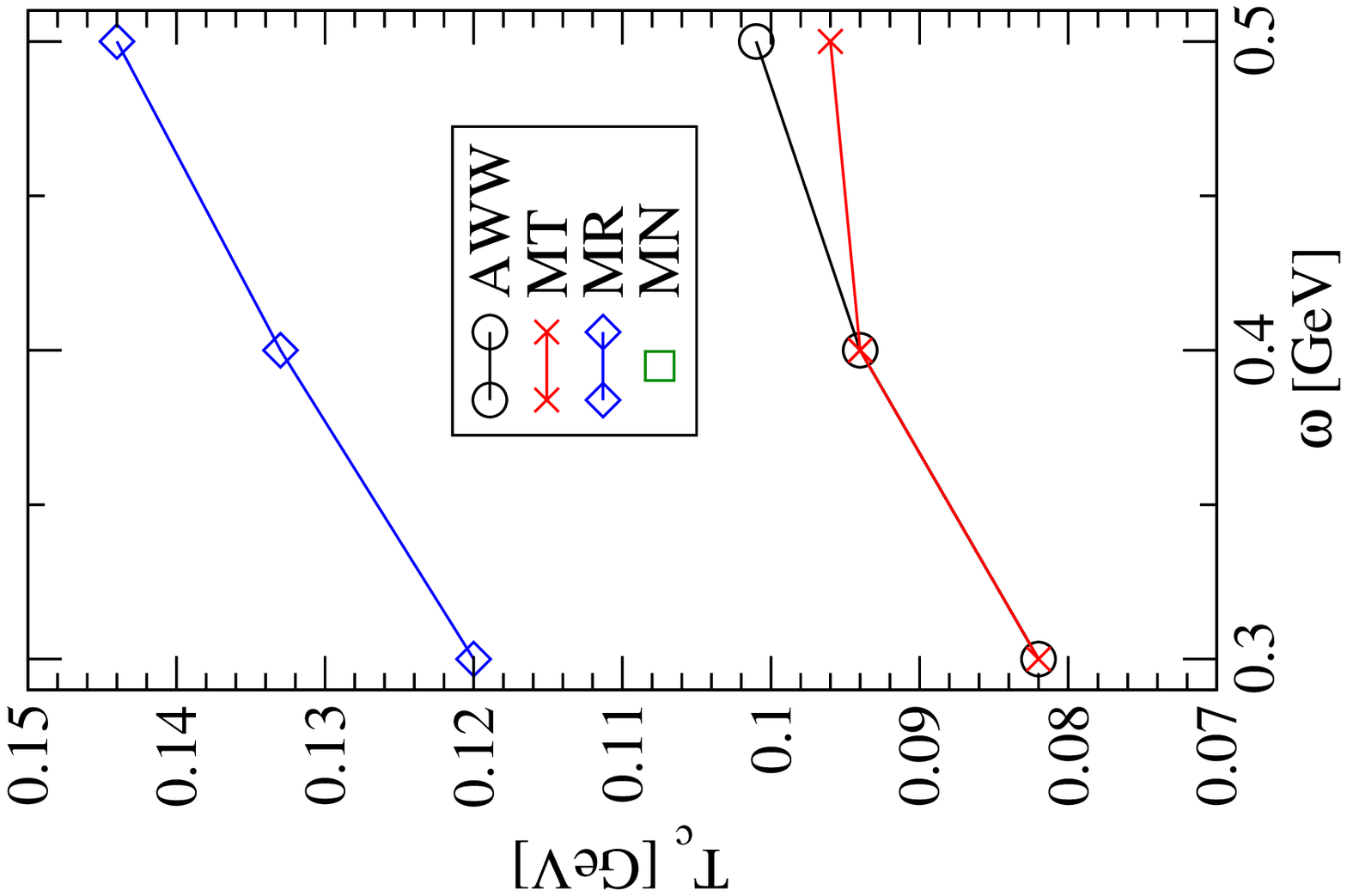}
\includegraphics[scale=0.37,angle=270,clip=true]{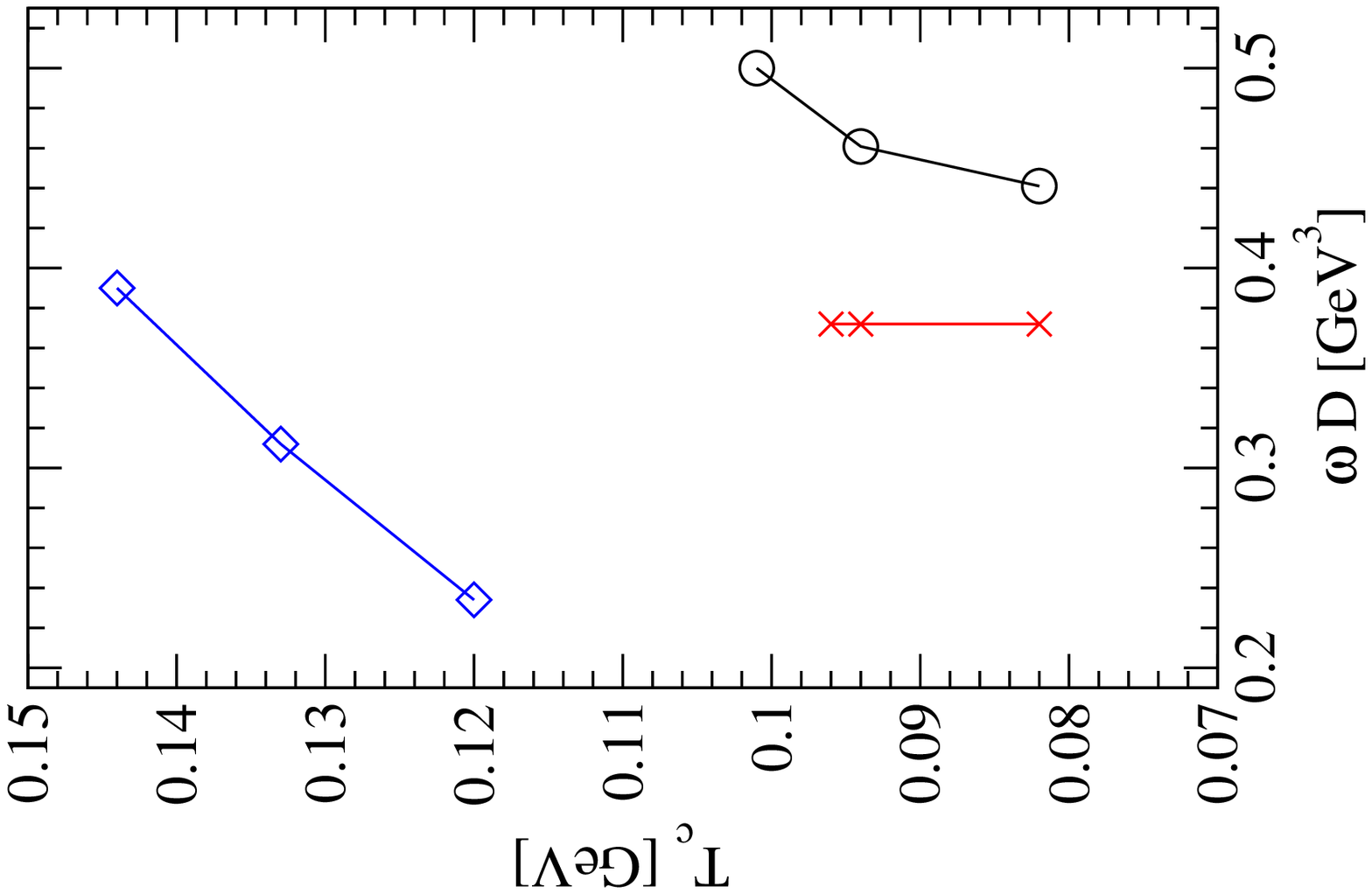}
\includegraphics[scale=0.37,angle=270,clip=true]{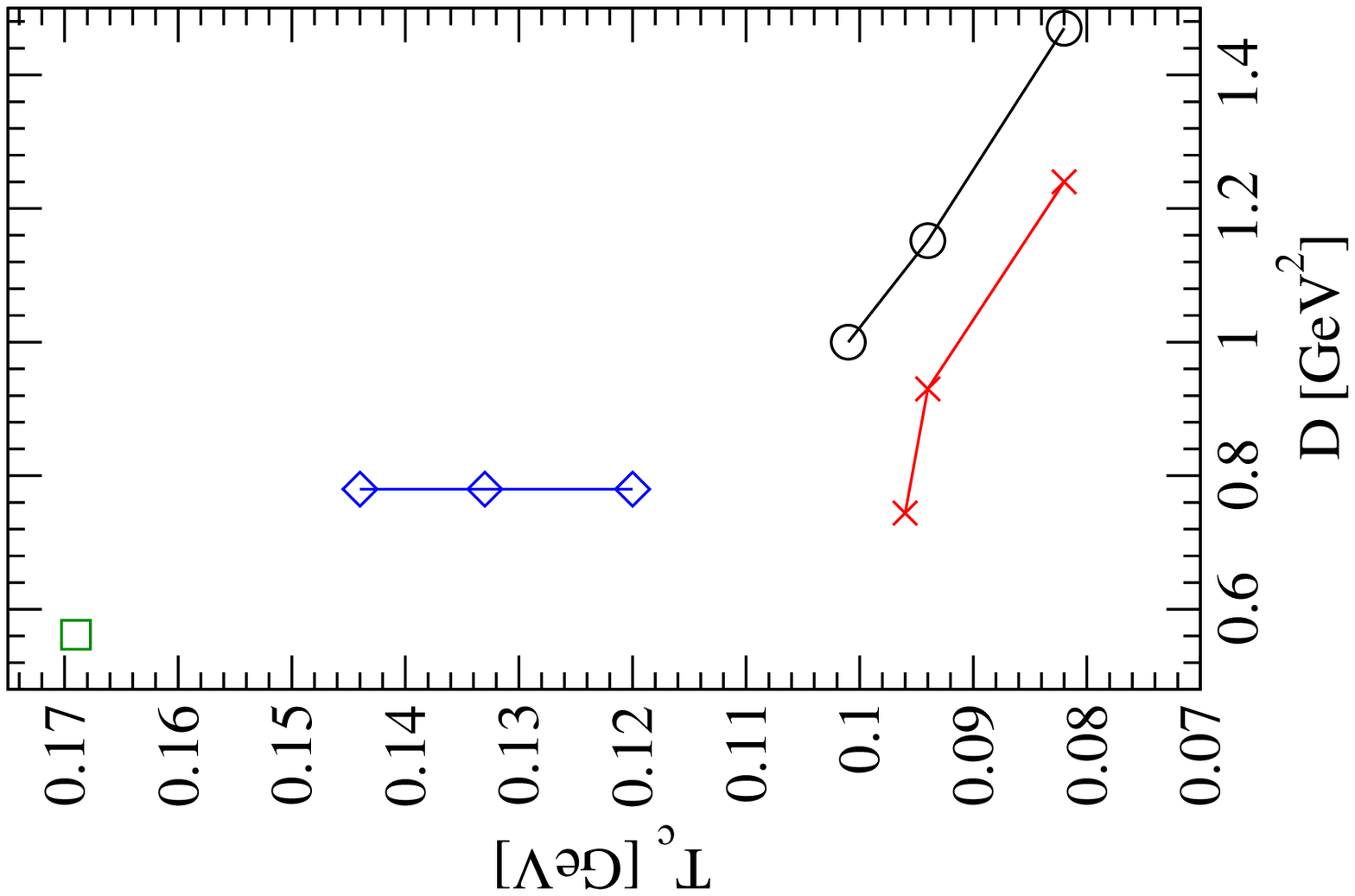}
\includegraphics[scale=0.37,angle=270,clip=true]{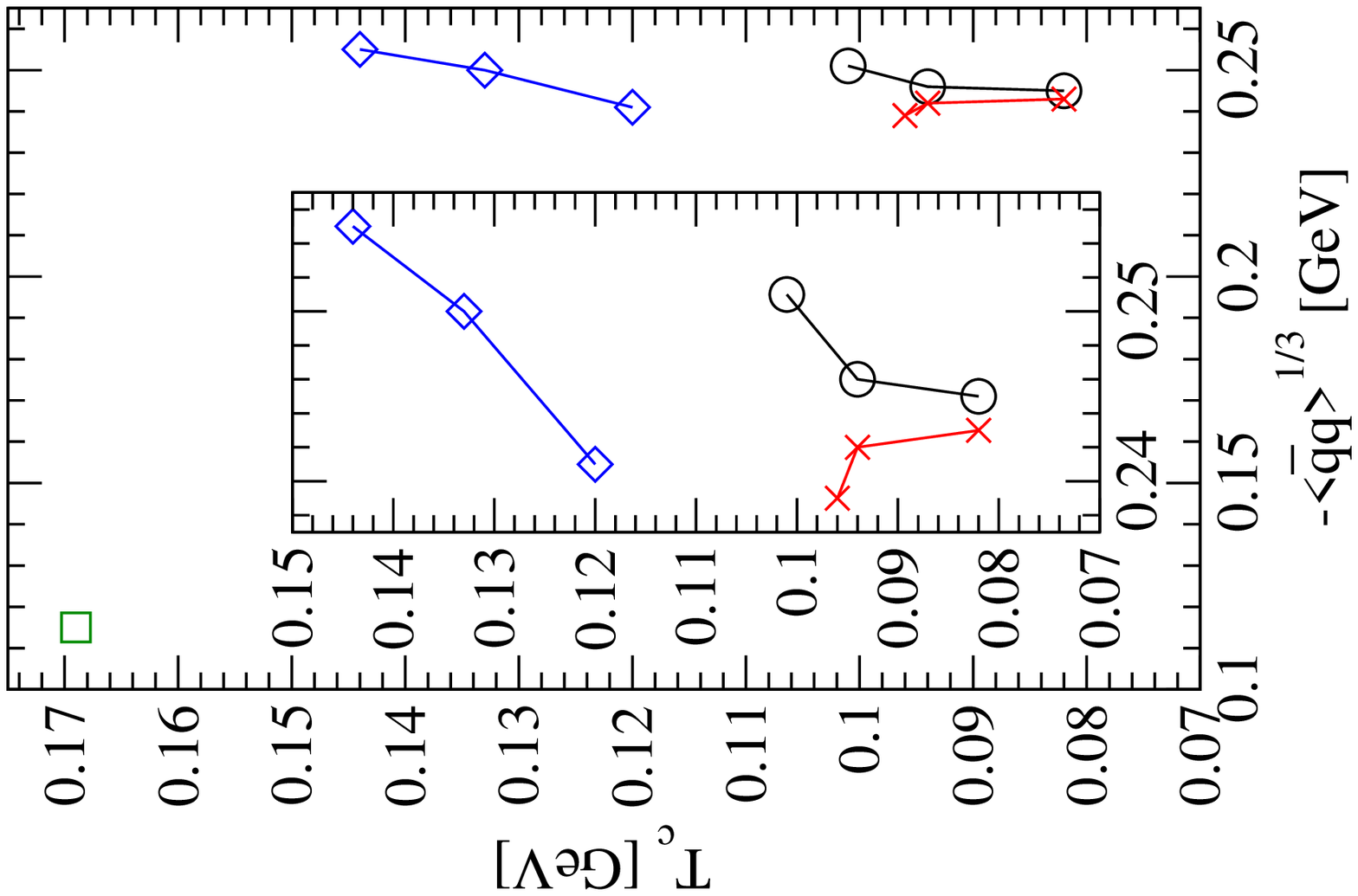}
\caption{(Color online) From left to right, the four panels show the results for $T_c$ as a
function of: $\omega$, leftmost panel (a); $\omega\,D$, left-middle panel (b); $D$, right-middle panel (c); and
the chiral condensate, rightmost panel (d). The different colors and symbols on the lines denote the
corresponding interactions, namely: black (circles) -- AWW; red (X) -- MT; and blue (diamonds) -- MR.
The result for MN (single green square) is only shown in panels (c) and (d), since it does not
have a value of $\omega$ associated with it. The insert in panel (d) enlarges the region of interest
for the $\omega$-dependent interactions.\label{fig:results}}
\end{figure*}

For a given interaction, $T_c$ increases with $\omega$, which is illustrated in panel (a) of
Fig.~\ref{fig:results}. A straight-forward interpretation of this observation goes back to
$\omega$ representing an inverse effective range of the interaction. In this picture, $T_c$ is
higher for an interaction, for which this range is smaller, i.e. the major strength in the
interaction comes from a region defined by the larger momentum scale $\omega$. However, this
effect is obviously minor compared to the differences with regard to the form of the interaction,
in particular the appearance of the $\delta$-function term.

To further illustrate the details of the differences between the various forms for the interaction
as well as to highlight their characteristic features, we plot $T_c$ as a function of three more
specific properties of each model, namely $\omega\,D$ in panel (b), $D$ in panel (c), and
the chiral condensate in panel (d). While the attempt to quantitatively correlate $T_c$ globally
(i.e.~across the different interactions) to any of these fails, a clear structure is visible. 
Throughout all panels of Fig.~\ref{fig:results} the AWW and MT results are rather close together, but
clearly separate from MR, which again is clearly separate from MN.

Both $\omega\,D$ and $D$ investigated in the two middle panels have been interpreted as an ``integrated
strength'' of the interaction. In fact, the latter is proportional to the integral over $d^4q$ of the
interaction at zero temperature for all interactions investigated here, if one leaves out the UV part
$\mathcal F_{UV}$ in Eqs.~(\ref{eq:mt}) and (\ref{eq:mr}).
Fig.~\ref{fig:results} in panel (c) clearly shows that there is no overall simple dependence of $T_c$ on
$D$. In the MT interaction, a constant value for $\omega\,D$ leads to (almost) unchanged masses and
decay constants for ground-state pseudoscalar and vector mesons, for which reason $\omega\,D$ can be termed
``integrated strength'' in this case instead of $D$. Again, Fig.~\ref{fig:results} in panel (b) indicates
no simple dependence of $T_c$ on $\omega\,D$. In both cases, the absence of a simple relation is
exemplified already by the fact that three points with the same $\omega\,D$, the characteristic feature for MT,
in Fig.~\ref{fig:results} (b) and three points with the same $D$, the characteristic feature for MR, in
Fig.~\ref{fig:results} (c) each correspond to three different values of $T_c$.

Finally, in Fig.~\ref{fig:results} (d) we plot $T_c$ as a function of an order parameter of chiral
symmetry breaking, the chiral condensate, evaluated at $T=0$. In a very simple picture, one could argue
that, given a certain form of the dependence of the order parameter on the temperature (see,
e.g.~Fig.~\ref{fig:orderpar}), an enlargement of the value of the order parameter at $T=0$ would
naturally lead to an increase in $T_c$. However, panel (d) clearly shows that the interactions used
in our setup are not that simple.

Overall, these observations lead to the important possibility to establish distinct ranges of
values for $T_c$ accessible for each type of interaction, which has the potential to rule out certain
forms in a given truncation. We note here that various effects need to be taken into account
beyond the simple setup presently used; we will attempt to quantify some of them below. At this
point, however, we can still try to attribute the clear differences apparent from all panels in
Fig.~\ref{fig:results} to the structure of the interaction. The object of interest in this
respect is the appearance and relative strength of the $\delta$-function term in $\mathcal{G}$, Eqs.~(\ref{eq:mn})
-- (\ref{eq:mr}): In the AWW and MT interactions, Eqs.~(\ref{eq:aww}) and (\ref{eq:mt}), there is no
such term. In the MR interaction, Eq.~(\ref{eq:mr}), it carries half the strength of the coupling
in the sense of the integral $\int d^4q \mathcal{G}\sim D$ described above. In the MN interaction,
Eq.~(\ref{eq:mn}), it is the only term and obviously represents all of the interaction's strength.
From this, one can say that a less pronounced $\delta$-function term in the interaction will lead to lower
values of $T_c$.

\section{Beyond rainbow truncation\label{brt}}
While the rainbow truncation of the gap equation provides a simple setup for computations,
one has to keep in mind that the necessary corrections in any given circumstance may
change its results considerably. Therefore we discuss here some of the most important effects
from which a change, in particular to the critical temperature, is to be expected.

From a phenomenological point of view one can try to quantify the contribution of the rainbow-truncated
gap equation to the full (untruncated) result. Such an attempt has been made for meson properties at zero
temperature in \cite{Eichmann:2008ae} and has been extended to the baryon sector and exemplified for
several hadronic observables in \cite{Nicmorus:2008vb}. The idea is to change the values of the available
model parameters such that the resulting hadron properties, e.g.~the rho-meson mass, are deliberately
overestimated by the rainbow-ladder truncation result. In this way one expects corrections beyond
this truncation to bring it to the experimental value.

While we do not want to discuss the particular
assumptions made in \cite{Eichmann:2008ae,Nicmorus:2008vb}, the effect described there is relevant
for our present study, since we start with model parameters fixed to meson phenomenology at zero
temperature. To get a first estimate of the effects from a change of parameters at $T=0$ on $T_c$, we
simply adapt the two parameter sets referred to as ``A'' (which corresponds to our MT2 parameter set) and
``B'' (which has increased strength in the interaction to overestimate the rho-meson mass as discussed above)
in Ref.~\cite{Nicmorus:2008vb} and add $T_c$ to the Tab.~I given there. In this table, a systematic ratio
of $\approx 0.74$ is found for value in A divided by value in B for several hadron properties (e.g.~$m_\rho$,
$f_\pi$, $m_N$, $m_\Delta$, etc.). Our values for $T_c$ in this respect are $94$ MeV for set A and
$129$ MeV for set B, which produces a ratio of $0.73$ and thus fits perfectly into the picture.

For the present discussion here the relevant point is that parameter changes
inferred from the phenomenological estimation of corrections expected beyond rainbow-ladder truncation
are in agreement with investigations at zero temperature. In particular, one obtains an increase of $T_c$ by
about one third, depending on the details of the corrections assumed.

A different investigation by D.~Blaschke, D.~Horvati\'{c}, and D.~Klabu\v{c}ar regarding the role of the
Polyakov loop in an analogous, separable setup of the quark DSE at finite temperature indicates
that the inclusion of a Polyakov-loop potential shifts $T_c$ up by $\approx 70$ MeV \cite{horvatic:pc}.
This is also observed in other approaches such as Polyakov-quark-meson studies or 
Polyakov-Nambu--Jona-Lasinio-type models, see e.g.~\cite{Schaefer:2007pw} and references therein.

Finally, we briefly quantify corrections from inclusion of more structure in the quark self-energy,
in particular in the dressed quark-gluon vertex.
Ideally, one would together with the gap equation also solve the Dyson-Schwinger equations
for the gluon propagator and the quark-gluon vertex---the two unknown ingredients in the
DSE for the quark propagator---as self-consistently as possible at any given finite temperature.
Such investigations have already been started at $T=0$ (see e.g.~\cite{Alkofer:2008tt} and references
therein) and represent a natural and very promising, but at the same time rather involved path for
future extensions of the present study.

In a more immediate way, one can identify corrections
to rainbow truncation from the terms in the quark-gluon-vertex DSE and attempt to incorporate
these into the gap equation. Pion-loop contributions have been investigated at $T=0$ in
\cite{Fischer:2007ze}. Ref.~\cite{Mueller:2008am} reports preliminary results of an analogous study
at finite temperature, which indicate a reduction of $T_c$ of about 5\% upon the inclusion of
pion-loop, but not scalar-meson-loop corrections in the quark self energy and no change of the 
(mean field) critical exponents.

Another effect, coming from the combination of two other vertex corrections, can be estimated by an
extension of a simple MN-study of quark and meson properties presented in \cite{Bhagwat:2004hn} to
finite temperature. This is a straight-forward procedure, which has many aspects
and will be investigated in full detail elsewhere \cite{jmaankinprep}. For the moment we note that
a first estimate of shifts in $T_c$ are of the order of 15\%. Another interesting observation is
that in the course of dressing the bare quark-gluon vertex one can arrive at results with the possibility
for solutions of the gap equation with $D(\vec{p}\,^2,\omega_k)\ne 0$.

To conclude this section, we remark that corrections to the rainbow truncation of the quark DSE should 
not be viewed as corrections to a mean-field approximation; corrections in the quark self energy in 
our present calculations happen at the level of the structure of the quark-gluon vertex. We would also like
to emphasize that, although not demonstrated explicitly here, for finite current-quark masses we obtain
a smooth crossover instead of a second-order phase transition.

\section{Conclusions and Outlook\label{conclusions}}
We have investigated the impact of various forms of an effective quark-gluon interaction in the rainbow-truncated
quark DSE in QCD at finite temperature on the critical temperature of the chiral phase transition.
The established mean-field behavior of this transition in the present truncation is confirmed throughout.
Regarding non-universal aspects of the transition, it is apparent that there is no simple overall relationship between
the distribution of the infrared strength of the interaction and the critical temperature, although 1) a trend
can be extracted identifying the part of the interaction strength carried by a $\delta$-function term in some
of the models as a qualitatively distincitve parameter, and 2) within the various models simple relations
of $T_c$ and relevant model parameters do exist.

Changes of $T_c$ due to various corrections to the present truncation are quantified and put into context.
In addition to effects already described in the literature we estimate particular types of corrections
in the present work using phenomenological arguments on one hand as well as a simple model-extension of the quark-gluon vertex
on the other hand. Alltogether, the various corrections put our results in a consistent perspective with
regard to the results for $T_c$ obtained in other approaches.

Furthermore, the present results clearly show that the chiral transition temperature is an excellent 
candidate for detailed studies of various aspects of the strong interaction. In particular it distinctively 
relates to both assumptions about as well as particular and detailed properties of the effective 
interaction between quarks and gluons in the DSE setup.
Further studies beyond the present investigation will shed more light on this relationship as well as
on how the mean-field behavior of the chiral phase transition may change beyond the current truncation.


\begin{acknowledgments}
The authors would like to acknowledge valuable discussions with R.~Alkofer, G.~Eichmann,
D.~Horvati\'{c}, D.~Klabu\v{c}ar, and B.-J.~Schaefer. A.K.~would like to thank his colleagues at
the Institute for Theoretical Physics of the University of Zagreb for their hospitality during
the initial phase of this work.
This work was supported by the Austrian Science Fund \emph{FWF} under project no.\ P20496-N16, the
Austrian Research Association \emph{\"OFG} under MOEL stipend no.\ 248, and
is performed in association with and supported in part by the \emph{FWF} doctoral program no.\ W1203-N08.
\end{acknowledgments}

%\bibliography{had_nucl_graz}

\end{document}